\documentclass[nofootinbib,a4paper,aps,pra,showpacs,preprintnumbers,twocolumn]{revtex4-2}
\usepackage{cmap}
\usepackage{graphics,graphicx,epsfig}
\usepackage{epstopdf}
\usepackage[centertags]{amsmath}
\usepackage{amsfonts}
\usepackage{amssymb}
\usepackage[cp1251]{inputenc}
\usepackage[english]{babel}
\inputencoding{cp1251}
\usepackage{graphicx}
\usepackage{amsthm,color}
\usepackage{euscript}
\graphicspath{{pics/}}
\DeclareGraphicsExtensions{.pdf,.png,.jpg,.eps}
\begin{document}
	\def\BY{\begin{eqnarray}}
		\def\EY{\end{eqnarray}}
	\def\L{\label}
	\def\nn{\nonumber}
	\def\ds{\displaystyle}
	\def\o{\overline}
	\def\({\left(}
	\def\){\right)}
	\def\[{\left[}
	\def\]{\right]}
	\def\<{\langle}
	\def\>{\rangle}
	\def\h{\hat}
	\def\td{\tilde}
	\def\r{\vec{r}}
	\def\ro{\vec{\rho}}
	\def\h{\hat}
	\def\v{\vec}
	\title{Broadband chip-based source of a quantum noise with electrically-controllable beam splitter}
	\author{E. A. Vashukevich$^1$, V. V. Lebedev$^{2,3}$, I.~V.~Ilichev$^{2,3}$, P. M. Agruzov$^{2,3}$, A. V. Shamrai$^{2,3}$, V. M. Petrov${^3}$, T. Yu. Golubeva$^1$,}
	
	\affiliation{$^1$ Saint Petersburg State University, Universitetskaya nab. 7/9, St. Petersburg, 199034 Russian Federation,\\
		$^2$ Ioffe Institute, Politekhnicheskaya str. 29, St. Petersburg, 194021 Russian Federation\\
	$^3$ ITMO University, Kronverksky pr. 49, St. Petersburg, 197101 Russian Federation}
	\begin{abstract}
For the first time, the theory and practical realization of a broadband quantum noise generator based on original integrated optical beam splitter in the form of a Mach-Zehnder interferometer is demonstrated. The beam splitter with a double output, made on a lithium niobate substrate, provided accurate electro-optical balancing of the homodyne quantum noise detection circuit. According to our knowledge, the experimentally obtained excess of quantum noise over classical noise by 12 dB in the frequency band over 4 GHz, which is the best parameters of quantum noise generators known from the literature.
	\end{abstract}
	\maketitle

\section{Introduction}

Quantum noise generators and random number generators based on them are on demand for many applications \cite{1,2,3}. The homodyne detection of vacuum fluctuations is one of the most efficient techniques to build a quantum noise generator. As a rule, quantum noise generators are used for the subsequent development of quantum random number generators. To do this, the analog signal must be converted into a digital code \cite{4,5,6,7,8,9,10}. 

Quantum vacuum fluctuations are used as a physical source of entropy of the noise generator. Its technical implementation is based on a local oscillator, beam splitter, and balanced detection scheme, that provides suppression of classical noise and registration of quantum shot noise. From the point of view of the informational throughput of the random number generator, one of the most important parameter here is the frequency band of the quantum noise at the output of the balanced detector \cite{6}. The currently experimentally achieved maximum band of homodyne detection of vacuum fluctuations is about 1 GHz \cite{8, 9}, which is due to the use of schemes on the so-called "volumetric" optics.

Integrated optical beam splitters based on silicon optical waveguides \cite{10} can only partially solve the problems of volumetric optics. Sufficiently high absorption and photosensitivity at telecommunications wavelengths (1500 - 1600 nm) produce sources of additional classical noise and limit the maximum optical power, and the thermoelectric control used in \cite{10} for active tuning has a high inertia and is not suitable for broadband devices. This did not allow the generation of quantum noise in the band of more than 150 MHz.

To construct a quantum noise generator, a scheme based on balanced homodyne detection of a vacuum field is widely used. However, such an implementation of the generator has a significant limitation: the lack of the ability to control interference when mixing fields on the beam splitter. Violation of the ideal symmetry of the beam splitter coefficients leads to a violation of the balance on the detectors, which negatively affects the visibility of the signal, and, as a result, the speed of random number generation. To solve this problem and implement the control of balanced detector, we used an integrated optical Mach-Zehnder interferometer formed by input Y-branch, output X-coupler, and with the electro-optical control of phase difference between arms of the interferometer.

Note that the Y-branch despite the presence of only one input port (Fig. \ref{Fig1}) serves as a mixer of fields of a local oscillator and vacuum fluctuations. Vacuum fluctuations penetrate in the Y-branch from substrate as leakage modes. It is not difficult to prove that there is a second port, given the unitarity of conversion of the input radiation to the output produced by the beam splitter. By organizing the illumination of the circuit from the output side and varying the phase difference of the fields, it is possible to see the points of the output radiation through the substrate in a situation where the central the mode is suppressed by destructive interference. It is these output points that will correspond to the second, unlit (vacuum) input of the beam splitter when the circuit is normally illuminated from left to right. Thus, we will describe the input Y-branch of the scheme under consideration as a four-port device \cite{11} like a X-coupler or volume beam splitter, on one of the ports of which there is a field in the vacuum state. Then the whole system under consideration is similar to the “usual” Mach-Zehnder interferometer with two inputs and two outputs.  
 
\section{Theory}
\begin{figure*}
	\includegraphics[scale=0.65]{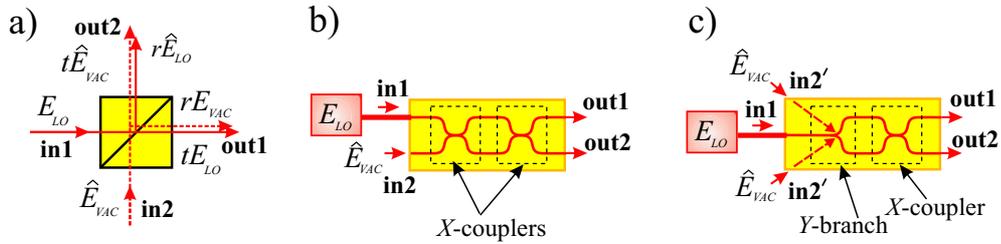}
	\caption{ a) Definition of the coefficients $r_i$ and $t_i$ for the volumetric beam splitter cube. b), c) the possible types of tunable integrated optical beam splitters. b) beam splitter, based on two X-couplers, that form the Mach-Zehnder interferometer with 2 inputs and 2 outputs; the  input 1 is directly connected by optical fiber to the source of $E_{LO}$,  only the mode of the local oscillator gets to this input; the input 2 is open, only the modes of vacuum fluctuations corresponding to the modes of the optical waveguide enters this input;  c) based on one Y-branch and one X-coupler, that form the Mach-Zehnder interferometer with 1 input and two outputs. The additional, second input in2’ for vacuum fluctuations occurs due to the leakage modes (dashed arrows).}\label{Fig1}
	\end{figure*}

Let us consider the case when the strong classic field from the local oscillator $E_{LO}(z,t)$ enters the beam splitter at the input 1, and only the vacuum field  $\h E_{vac}(z,t)$ enters at the input 2 (Fig. \ref{Fig1}). After mixing on the first beam splitter, the fields are given a relative phase delay $\phi$, after which the fields are mixed again on the second beam splitter and detected. It is the phase difference $\phi$ that acts as a parameter that additionally controls the interference conditions on the second beam splitter. The transformation of the fields on the first and second beam splitters can be set by matrices  $M_{BS,1}, M_{BS,2}$:
\BY
M_{BS,i}=\begin{pmatrix}\cos{\(\alpha_i\)}&\sin{\(\alpha_i\)}\\\sin{\(\alpha_i\)}&-\cos{\(\alpha_i\)}\end{pmatrix},\;\;i=1,2.
\EY
Here the parameters $\alpha_1,\alpha_2 $ are set so the $\cos{\(\alpha_i\)}$ is equal to the amplitude transmission coefficient of the beam splitter $t_i$, and the $\sin{\(\alpha_i\)}$  is correspondingly equal to the reflection coefficient $r_i$. 
It can be noted that with the chosen parametrization, the conservation law $r_i^2+t_i^2=1$ is fulfilled automatically.

The phase delay in one of the arms of the interferometer is given by the matrix:
\BY
M_{Ph}=\begin{pmatrix}\exp{\{i \phi/2\}}&0\\0&\exp{\{-i \phi/2\}}\end{pmatrix}
\EY
Next, one can write down the expression for the fields at the output of the scheme $\h E_{out,1},\h E_{out,2} $ in terms of the initial fields:
 \BY
 &&\binom{\h E_{out,1}}{\h E_{out,2}}=U\binom{ E_{LO}}{\h E_{vac}}\\
 &&U=M_{BS,2}\cdot M_{Ph}\cdot M_{BS,1}\label{4}
 \EY
Here the elements of the of the transformation matrix are:
\BY
 &&U_{11}=e^{i\phi/2}\(\cos{\(\alpha_1\)}\cos{\(\alpha_2\)}+e^{-i \phi}\sin{\(\alpha_1\)}\sin{\(\alpha_2\)}\)\\
&&U_{12}=e^{i\phi/2}\(\sin{\(\alpha_1\)}\cos{\(\alpha_2\)}-e^{-i\phi}\cos{\(\alpha_1\)}\sin{\(\alpha_2\)}\)\;\;\;\;\;\;\\
&&U_{21}=e^{i\phi/2}\(\cos{\(\alpha_1\)}\sin{\(\alpha_2\)}-e^{-i\phi}\sin{\(\alpha_1\)}\cos{\(\alpha_2\)}\)\;\;\;\;\;\;\\
&&U_{22}=e^{-i\phi/2}\(\cos{\(\alpha_1\)}\cos{\(\alpha_2\)}+e^{i\phi}\sin{\(\alpha_1\)}\sin{\(\alpha_2\)}\)\;\;\;\;\;\;
\EY
The photocurrent operators on both detectors can be written as follows:
\BY
&&\h j_1= \h E^\dag_{out,1}\h E_{out,1}=(U^*_{11}E^*_{LO}+U^*_{12}\h E^\dag_{vac})\nn\\&&\times(U_{11}E_{LO}+U_{12}\h E_{vac})\\
&&\h j_2= \h E^\dag_{out,2}\h E_{out,2}=(U^*_{21}E^*_{LO}+U^*_{22}\h E^\dag_{vac})\nn\\&&\times(U_{21}E_{LO}+U_{22}\h E_{vac})
\EY
Now, for convenience, let’s move on to the quadrature components (with numeric representation for the classical field and operator for the quantum one): $E_{LO}=\varepsilon^\prime+ i\varepsilon^{\prime\prime}$, $\h E_{vac}=\h x+ i \h y$; in addition, let us denote $|E_{LO}|^2=I_{LO}$. The differential signal $\h j_-(\phi)$ can be written as follows (we denote $\alpha_\pm=2(\alpha_1\pm\alpha_2)$):
\BY&&\h j_-(\phi)=\h i_1-\h i_2=\left[I_{LO}-\h x^2-\h y^2\right]\nn\\
&&\times\left[\cos^2(\phi/2)\cos(\alpha_-)+\sin^2(\phi/2)\cos(\alpha_+)\right]\nn\\
&&+2\left[\varepsilon^\prime\h x+\varepsilon^{\prime\prime}\h y\right]\nn\\
&&\times\left[\cos^2(\phi/2)\sin(\alpha_-)+\sin^2(\phi/2)\sin(\alpha_+)\right]\nn\\&&-2 \left[\varepsilon^\prime\h y-\varepsilon^{\prime\prime}\h x\right]\sin (\phi) \sin (2 \alpha_2)
\EY

In the case when the beam splitters are symmetric $r_i=1/\sqrt{2}=t_i$, the transformation matrix has a simple form:
\BY
U=\begin{pmatrix}\cos{\phi/2}&i \sin{\phi/2}\\i \sin{\phi/2}&\cos{\phi/2}\end{pmatrix}
\EY
The difference photocurrent with symmetric beam splitters  $j^{s}_-(\phi)$ will be modulated by the phase difference $\phi$:
\BY
&&\h j^{s}_-(\phi)=(I_{LO} -\h x^2 - \h y^2)\cos{\phi} -2 (\varepsilon^\prime \h y -\varepsilon^{\prime\prime}\h x)\sin{\phi}\;\;\;\;
\EY
As you can see, with the phase difference $\phi=\pi/2$ the first term proportional to the intensity of the fields will be completely suppressed, and homodyne detection of the quadrature components of the quantum field is carried out in the Mach-Zehnder interferometer scheme. For simplicity, here and further, we will select the phase of the local oscillator so that $\varepsilon^{\prime\prime}=0$, then the difference signal can be written as follows:
\BY
\h j^{s}_-(\pi/2)=-2\varepsilon^\prime \h y \L{9}
\EY

Since the value of the quadrature component measured in the experiment is a random variable, we have the opportunity to generate a sequence of truly random numbers in the proposed scheme. As can be seen from (\ref{9}), the dispersion of the noise field y increases in proportion to the magnitude of the field of the local oscillator.

Now we will take into account the differences between the real experimental situation and the ideal one described above. First of all, we take into account the possible asymmetry of the beam splitter. Let’s assume that one of the beam splitters (for example, the output one) is asymmetric, choosing $r_2=\sqrt{0.49}, t_2=\sqrt{0.51}$. Now we write down the dependence of the difference current in the case of an asymmetric beam splitter  $\h j^{a}_-(\phi)$ on the phase, leaving the field of the local oscillator purely real:
\BY
\h j^{a}_-(\phi)=&& 0.9998 \cos (\phi) \left(I_{LO}^2-\h x^2-\h y^2\right)\nn\\&&-2\cdot0.9998\sin (\phi) \varepsilon^\prime \h y-2\cdot0.02\varepsilon^\prime \h x \L{10}
\EY

As one can see, the second quadrature component begins to appear in the signal. Moreover, the last term in (\ref{10}) is phase-independent. However, the contribution from the first term, proportional to the intensity of the fields, which is most harmful for noise generation, can be completely compensated by choosing a suitable phase of the modulator. When the phase difference is $\phi=\frac{\pi}{2}$ we will get:
\BY
&&\h j^{a}_-(\frac{\pi}{2})=-2\cdot0.9998 \varepsilon^\prime \h y-2\cdot0.02\varepsilon^\prime \h x
\EY

Note that the obtained form of recording the signal allows us to talk about the measurement of the generalized quadrature of the noise field, that is, the measurement in the basis expanded by some angle. Since the distribution of the vacuum field on the phase plane is absolutely symmetric, the reversal of the basis does not introduce any changes in the operation of the noise generator. Thus, in the absence of other factors of the “imperfection” of the scheme, the asymmetry of the beam splitter would be insignificant for the generation of random numbers. 

The key factor here, however, is the fact that if the beam splitter is not symmetric, an additional classical noise component will be present in the difference current, which is completely subtracted when considering a symmetric circuit. We show this explicitly by introducing losses  $\eta_1, \eta_2$, in both arms of the interferometer associated with the presence of classical noise. Then, instead of the Eq. (\ref{4}) used above, the field at the output of the interferometer will be set by the expression:
\BY
 &&\binom{\h E_{out,1}}{\h E_{out,2}}=\tilde U\binom{ E_{LO}}{\h E_{vac}}\\
&&\tilde U=M_{BS,2}\cdot\begin{pmatrix}\eta_1&0\\0&\eta_2\end{pmatrix} M_{Ph}\cdot M_{BS,1}
\EY

We repeat all the calculations made without taking into account losses, preserving the assumptions made earlier about the symmetry of the first beam splitter and the realness of the field of the local oscillator. The difference current $\h j_-(\phi)$ will then have the form:
\BY
&&\h j_-(\phi)=\frac{1}{2} \cos(2\alpha_2) \left(\eta_1^2-\eta_2^2\right) \left(I_{LO}+\h x^2+\h y^2\right)\nn\\&&+\eta_1\eta_2 \sin(2\alpha_2)\cos(\phi)\left(I_{LO}-\h x^2-\h y^2\right)\nonumber\\
&&+\varepsilon^\prime \h  x \cos (2\alpha_2) \left(\eta_1^2+\eta_2^2\right)-2\eta_1\eta_2\varepsilon^\prime \h y\sin (2 \alpha_2) \sin (\phi )\;\;\;\;\label{21}
\EY

In the case of symmetrical output beam splitter  $\alpha_2=\pi/4$, the expression (\ref{21}) is reduced by the choice of phase $\phi=\pi/2$ to (\ref{9}) with extra multiplication by factor $\eta_1\eta_2$. However, as one can see, for any asymmetric beam splitter in (\ref{21}), there remains a phase-independent first term containing the intensity of the field of the local oscillator, which will dominate the signal of interest. This term can be removed by analyzing the phase $\phi$ and selecting it so that the following equation is performed:
\BY
\cos(\phi)\approx\frac{\left(\eta_1^2-\eta_2^2\right)}{2\eta_1\eta_2}\cot(2\alpha_2)
\EY

Such a choice of the modulator phase leads to complete mutual compensation of the first two terms in expression (\ref{21}). As a result, the expression for the difference photocurrent can again be represented as the amplification of the generalized noise quadrature (similar to Eq. (\ref{10})), where the gain is lesser than the original one by a factor of $\eta_1\eta_2$. It is interesting to estimate the amount of phase adjustment required for balancing the circuit. For example, for $\cos(2\alpha_2)=-0.02, \eta_1=0.9,\eta_2=0.85 $ to compensate for the term containing $I_{LO}$, we need to add to the phase $\phi=\pi/2$ only the  $3\times10^{-4}\pi$. Thus, the use of a phase modulator makes it possible to balance the circuit in the presence of not only asymmetric detector operation, but also various classical noises in the interferometer channels.

\section{Experiment}
We have proposed and experimentally implemented a broadband quantum noise generator using an integrated optical Mach-Zehnder interferometer with a single input and a double output as an electrically-controlled beam splitter (BS), made on the basis of optical waveguides in a lithium niobate crystal substrate (Fig. 2). The $\hbox{LiNbO}_3$ congruent single crystal plate of the X-cut had the size $5\times50\times1$mm$^3$. The single-mode channel optical waveguides were manufactured using the technology of thermal diffusion of Ti-ions \cite{12}. Light propagated along Y crystallographic axes. A push-pull electrodes was deposited along one of the arms of the interferometer, which made it possible to adjust the amplitude transmission coefficient of the beam splitter $t_i$ using the electro-optical effect.
\begin{figure}
	\includegraphics[scale=0.45]{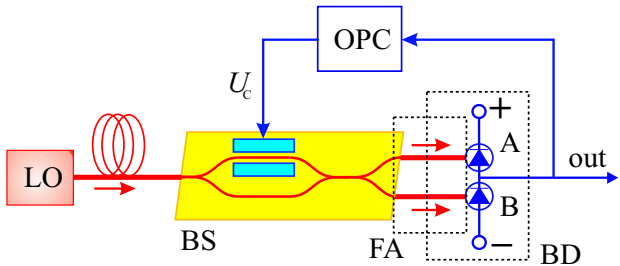}
	\caption{ The experimental realization: quantum noise generator based on integrated-optic Mach-Zehnder beam splitter with the electrical control. LO is the local oscillator, BS is the beam splitter, FA is the fiber-optic assembly, BD is the balanced detector, OPS is the operating point control system. 
	}\label{Fig2}
\end{figure}

A single frequency laser with distributed feedback with a wavelength of 1552 nm, a radiation line width of 170 kHz, and a power of 100 mW was used as a local oscillator (LO).

A special attention was paid to the design of the balanced detector (BD). The radiation from the beam splitter outputs was transmitted through the fiber assembly (FA) to the InGaAs-pin photodiodes (A,B) that form the balanced detector. The difference of the
optical paths of the fiber-optic assembly did not exceed 0.1 of the operating wavelength. 

The band of each photodiode was 10 GHz, which provided the band of the balanced detector above 4 GHz. For this purpose, photodiodes with the closest possible frequency response and the same sensitivity of  $\approx$0.78\!A/W were selected. A high saturation current ($\sim$ 30 mA) and a dark current of less than 1 $\mu$A provided a high dynamic range. The operating point control system (OPC) provided accurate balancing of output currents ($<$0.1$\%$) \cite{13}. To suppress classical noise, anti-phase subtraction of synchronous signals was provided by equalizing the optical and electrical paths in the balanced circuit. 

The efficiency of the balanced photodetector was evaluated by suppressing common-mode interference. To emulate common-mode signal the laser radiation was modulated in amplitude and differencial signal was applied to the tunable integrated optical BS. The suppression was defined as the ratio of the frequency response to a differential signal to the frequency response to a common-mode signal. The common-mode interference suppression by more than 15 dB is observed in the band over 3 GHz. The decrease in the suppression efficiency with increasing frequency was due to increased requirements for the accuracy of performing phase matching and the difference in the frequency response of photodiodes.

\begin{figure}[!htb]
	\includegraphics[scale=0.6]{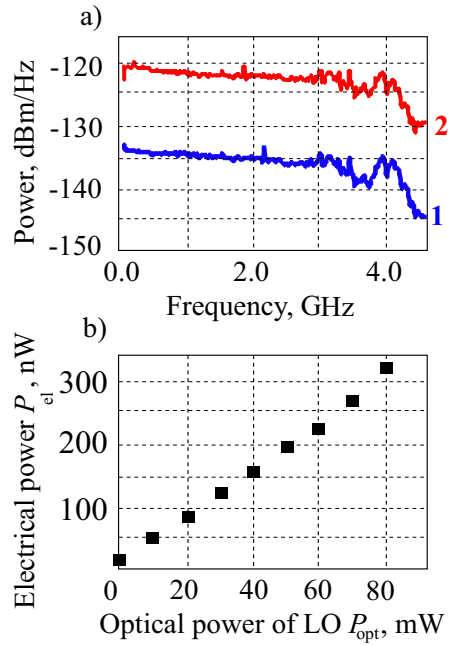}
	\caption{ Experimental results. a): the spectral power density  $N(f)$ at the output of balanced detector. 
		1: local oscillator “OFF”, 2: local oscillator “ON”; 
		b): the electrical power $P_{el}$ at the output of the balanced detector as a function of optical power $P_{opt}$ of the local oscillator (LO).
	}\label{Fig3}
\end{figure}

A part of the electrical signal from the output of the balanced detector was sent to  OPC unit, which generated a feedback signal a control voltage $\pm U_C$. This voltage was applied to  electrodes, which made it possible to change the phase delay between the arms, and, consequently, to control the splitting coefficient of the beam splitter. The accuracy of the splitting control was not worse as 0.1$\%$ in power.

Measurements of the noise signal at the output showed an excess of the spectral power density of the detected quantum noise by an amount of more than 12 dB above the level of technical noise of the measuring system with a preamp in the band of more than 4 GHz (Fig. \ref{Fig3}, a). The level of classical noise caused by random intensity noise (RIN) of the laser has a value much smaller than the quantum ones, which is confirmed by the proximity to the linear dependence characteristic of quantum shot noise (Fig. \ref{Fig3}, b). The power level of the recorded quantum noise is in good agreement with the theoretical estimation, which predicts a linear growth:
\BY
N(f)=2qP_{opt}AR_0|H^2_{pd}(f)|,
\EY
where $N(f)$ is the spectral power density, $q$ is the electron charge, $P_{opt}$ is the power on the input of the photodiodes, $A$ is the direct current sensitivity of the photodiodes, $H_{pd}(f)$ is the transfer function of the balanced photodetector, $R_0$ is the output loading resistor of the balanced detector.

According to the estimates obtained from the literature, we have developed a broadband quantum noise generator, probably with the highest parameters for the present time. Experimentally, an excess of quantum noise over classical noise was obtained by more than 12 dB in the band of more than 3 GHz, which is the best characteristics for this type of generators known from the literature. This generator is based on homodyne detection of quantum fluctuations, using a controlled beam splitter based on a Mach-Zehnder waveguide interferometer on a lithium niobate substrate and a high-frequency balanced detector.

\section{Conclusion}

We have designed and experimentally implemented a quantum noise source with remarkable characteristics. Three factors mainly determine the broadband source with a spectral bandwidth of more than 3 GHz.

First of all, such a broad generation band turns out to be possible due to the integrated-optical chip-based implementation of homodyne detection with the field mixer in the form of Y-branch. It should be noted that, in contrast to traditional volumetric beam splitters or waveguide X-couplers, the reflection and transmission coefficients of which depend quite critically on the the light wavelengtht \cite{15}, integrated optical Y-branches are much more broadband elements. Thus, the spectral width of our generator is determined by the spectral characteristics of the detectors, not the beam splitters. This factor allowed us to build a theory without considering the spectral dependence of the beam splitter coefficients.

The second important factor that makes it possible to achieve record results is developing a feedback loop that allows us to quickly and accurately control the delay in the arm of the Mach-Zehnder interferometer. As shown in the theoretical part of the paper, fine tuning of the balance allows getting rid of the influence of classical noises in the system. The presence of asymmetric losses in the interferometer channels leads to degeneration of interference and the uncompensated currents proportional to the homodyne intensity mixing into the signal. Even a weak asymmetry can significantly impair the signal visibility due to large values of the homodyne amplitude. Controlling the interference phase using a feedback loop enabled the achievement of the visibility of the quantum noise signal above the classical noise by more than 12 dB.

It should be noted that if we imagine an ideal situation in which there are no classical noises in the system, then the asymmetry of the beam splitter does not worsen the observation parameters of quantum noise but only rotates the observation basis, which is not significant for symmetric noise distribution. Correctness of the quantum-mechanical description of the measurement in the presence of a feedback loop should also be discussed here. It is a well-known situation when feedback stabilizes the photoelectron flux but degrades the noise characteristics of light that produce this flux \cite{16,17,18}. In the mentioned case, the direct connection between the operators of the light's quadratures and the photocurrent's operator is lost. There is no such problem in our case since the feedback controls not the quantum system but the classical one.

Finally, the third factor is the experimental selection of detectors with the closest possible characteristics. The selection of suitable photodiodes is an essential factor in improving the performance of the circuit, and we suggest that it is this block of the circuit that define cut-of frequency of the spectral characteristics of the device as a whole.


\begin{thebibliography}{100}%
	\bibitem{1} M. Herrero-Collantes, J. C. Garcia-Escartin, Rev. Mod. Phys. 89, 015004 (2017), DOI: 10.1103/RevMOdPhys.89.015004
	\bibitem{2}	A. V. Gleim, V. V. Chistyakov, O. I. Bannik, J. of Opt. Tech., 84 (6), 362 (2017). DOI: 10.1364/JOT.84.000362
	\bibitem{3}	Y. Liu, Q. Zhao, M. H. Li, Nature 562, 548–551 (2018). DOI: 10.1038/s41586-018-0559-3
	\bibitem{4}	C. Gabriel, C. Wittmann, D. Sych, Nature Photon 4, 711–715 (2010). DOI: 10.1038/nphoton.2010.197
	\bibitem{5} T. Symul, S. M. Assad, P. K. Lam, Appl/ Phys. Lett., 98, 231103 (2011). DOI: 10.1063/1.3597793
	\bibitem{6} T. Gehring, C. Lupo, A. Kordts, A. et al. Nat. Comm. 12, 605 (2021). DOI: 10.1038/s41467-020-20813-w
	\bibitem{7} H. Zhou, P. Zeng, M. Razavi, X. Ma, Phys. Rev. A, 98 (4) 042321 (2018). DOI: 10.1103/PhysRevA.98.042321
	\bibitem{8}	J. Y. Haw, S. M. Assad, A. M. Lance, N. H. Y. Ng, V. Sharma, P. K. Lam, T. Symul, Phis. Rev. Appl., 3, 054004 (2015). DOI: 10.1103/PhysRevApplied.3.054004
    \bibitem{9} B. Xu, Z. Chen, Z. Li, J. Yang, Q. Su, W. Huang, Y. Zhang, H. Guo, Quant. Sci. and Tech., 4 (2), 025013 (2019). DOI: 10.1088/2058-9565/ab0fd9
	\bibitem{10} L. Huang, H. Zhou, JOSA B, 36 (3), B130 (2019).DOI:10.1364/JOSAB.36.00B130
	\bibitem{11} M. Izutsu, Y. Nakai, and T. Sueta,  Opt. Lett. 7, 136-138 (1982). DOI: 10.1364/OL.7.000136
	\bibitem{12} O. Alibart, V. D'Auria, M. De Micheli, F. Doutre, F. Kaiser, L. Labonte, T. Lunghi, E. Picholle, and S. Tanzilli, J. Opt. 18, 104001 (2016), DOI: 10.1088/2040-8978/18/10/104001

	\bibitem{13} A. Petrov, A. Tronev, P. Agruzov, A. Shamrai, V. Sorotsky, Electronics. 9(11), 1861 (2020). DOI:~10.3390/electronics9111861
	\bibitem{15} S. M. Barnett,J. Jeffers, A. Gatti and R. Loudon,  Physical Review A, 57(3), 2134 (1998). DOI: 10.1103/PhysRevA.57.2134
	\bibitem{16} T. Golubeva, Yu. Golubev and D. Ivanov, Physical Review A, 75(2), 023815 (2007). DOI:~10.1103/PhysRevA.75.023815
		\bibitem{17} A. Masalov, A. Putilin and M. Vasilyev,  Journal of Modern Optics, 41(10), 1941-1953 (1994). DOI:~10.1080/09500349414551841
	\bibitem{18} H. Wiseman and G. Milburn, Physical Review A, 49(2), 1350 (1994). DOI:~10.1103/PhysRevA.49.1350
	\end{thebibliography}
\end{document}